\documentclass[11pt]{article}
\usepackage{aaspp4}
\usepackage{graphicx}
\newcommand\etal{{\it et al.}}
\newcommand{\NH}{\mbox{${\rm N}_{\rm H}$}}        

\begin{document}

\newlength{\bigpicsize}
\setlength{\bigpicsize}{3.5in}
\newlength{\smpicsize}
\setlength{\smpicsize}{2.2in}

\title{Soft X-ray Scattering and Halos from Dust}
\author{Randall K. Smith\altaffilmark{1} and Eli Dwek}
\affil{Code 685 \\ NASA Goddard Space Flight Center \\ Greenbelt, MD
20771}
\altaffiltext{1}{present address: Smithsonian Astrophysical
Observatory, MS 70, 60 Garden Street, Cambridge, MA 02138; email:
rsmith@cfa.harvard.edu}

\begin{abstract}

Small angle scatterings of X-rays by interstellar dust particles create
halos around x-ray sources. The halo intensity and its projected
radial distribution around a source can provide important
information on the spatial distribution of the dust along the line of
sight to the source, and on the physical properties of the scattering
dust particles.  Halos around x-ray point sources have been used by
several authors to infer that the scattering dust particles are
fluffy in nature, consisting of aggregates of smaller refractory
particles with vacuum occupying a significant fraction of their
volume.

The nature and morphology of interstellar dust particles has recently
gained new importance, since fluffy, composite dust particles have
been suggested as a possible solution to the interstellar carbon
``crisis.'' This crisis results from the discrepancy between the abundance
of carbon in the interstellar medium available for creating dust, and
the significantly larger amount of carbon that must be in dust in
order to account for the UV-optical interstellar extinction in the
diffuse ISM.

Previous studies of x-ray scattering have used the Rayleigh-Gans (RG)
approximation to the differential scattering cross section for
calculating halo properties.  However, the validity of the RG
approximation fails for energies below 1 keV.  We use the exact Mie
solution for the differential scattering cross section and find that
for these energies the scattering becomes much less efficient than is
predicted by the RG approximation.  Furthermore, the effects of K and
L band absorption by atoms in the dust become important.  The net
effect is that the RG approximation systematically and substantially
overestimates the intensity of the halo below 1 keV, relative to the
Mie solution result.  In particular Mathis \etal\ (1995) used the
weaker-than-expected halo intensity observed around Nova Cygni 1992 to
conclude that interstellar dust must be fluffy.  Using the Mie
solution to the scattering intensity and including the effects of
absorption, we find that, contrary to the conclusion of Mathis \etal\ 
(1995), the halo around Nova Cygni 1992 does not {\it require}
interstellar dust grains to be fluffy in nature, and that the data is
consistent with scattering from a mix of bare refractory silicate and
carbon grains as well.

\end{abstract}


\received{\underline{                                 }}
\revised{\underline{                                 }}
\accepted{\underline{                                 }}

\section{Introduction}

Measuring the small-angle scattering of X-rays from particles embedded
in a uniform medium to calculate their size and shape has long been a
standard technique in microbiology (Glatter \&
Kratky,\markcite{Glatter} 1982).  The scattering of X-rays from dust
grains in the interstellar medium (ISM) was first considered by
Overbeck\markcite{Overbeck} in 1965.  However, the weak x-ray halo
generated by small-angle x-ray scattering from dust has only recently
become observable (Mauche \& Gorenstein,\markcite{MG89} 1989; Mathis
\etal,\markcite{Mathis} 1995; Predehl \& Klose,\markcite{PK96} 1996).
In the laboratory, small-angle x-ray scattering observations can
measure the size and shape of the scattering particles quite
accurately, but astrophysical applications are still sharply limited
by the availability of sufficiently bright x-ray sources and the
spatial and energy resolution of x-ray telescopes.  However, recent
observations using the {\it ROSAT} Position Sensitive Proportional Counter
(PSPC) (Predehl \& Schmitt,\markcite{PS95} 1995) have found x-ray
halos around 25 point sources and 4 supernova remnants, using the
moderate $25''$\ resolution of this instrument.  The Advanced X-ray
Astrophysical Facility (AXAF) telescope will have $\sim 1 ''$\
resolution (Canizares, 1990), and should dramatically increase the
number of sources with visible halos.

In general the halo formed around an x-ray point source by small-angle
scattering depends, in addition to the spatial distribution of the
dust, upon many details of the interstellar dust grain population.
The intensity varies linearly with the number of grains along the line
of sight, and is highly dependent upon the size distribution of the
larger grains, which more efficiently scatter X-rays.  An individual
spherical dust grain scatters X-rays with a angular distribution
determined by the grain size distribution, composition, and density.
Given the dust parameters, the spatial distribution of the grains can
be estimated by comparing this scattered distribution with the
observed halo.  Although it is difficult to extract each dust grain
parameter separately, measurements of the halo's intensity as a
function of energy and angle from the source can provide useful limits
for the allowable dust grain model parameters.

The general theory of x-ray scattering by dust grains has been
discussed in detail by Mauche \& Gorenstein \markcite{MG86} (1986)
(MG86) and Mathis \& Lee\markcite{MathisLee} (1991) (ML91).  Their
results were based on the Rayleigh-Gans (RG) approximation to the
differential scattering cross section.  Predehl \&
Klose\markcite{PK96} (1996)(PK96) discussed the possibilities for
extracting dust grain parameters by observing x-ray halos with current
x-ray telescopes.  They also considered the limitations of the
Rayleigh-Gans approximation and the effects of absorption in the dust
grain on the scattered X-rays.  Recent observations of x-ray halos
(Mathis \etal\markcite{Mathis}, 1995; Woo \etal\markcite{Woo}, 1994;
and Predehl\markcite{Predehl}, 1996), using both the {\it ROSAT} and
{\it ASCA} telescopes, have found that the observed halo intensity is
less than that predicted by the general theory using the RG
approximation.  These authors concluded that their results were
evidence for ``fluffy'' dust grains which contain many holes and
voids.

These results are of particular interest because of the carbon
``crisis'' in the interstellar medium (ISM) (Cardelli
\etal\markcite{Cardelli}, 1996).  Observations have suggested that the
amount of carbon available to be in dust grains is substantially less
than the amount necessary to make the grains which account for the
observed extinction.  Mathis\markcite{Mathis96} (1996; M96) has
suggested that fluffy dust grains would account for the extinction
measurements while reducing the amount of carbon necessary to be
locked up in the dust.  

However, Dwek\markcite{Dwek97} (1997) pointed out that the composite
fluffy dust model (CFD) suggested by Mathis fails to resolve the
interstellar carbon crisis for several reasons: (1) the visual albedo
of the CFD particles is lower than the observed one. The increased
absorptivity of the composite dust grains produces therefore an excess
of infrared emission over that observed with the Diffuse Infrared
Background Experiment (DIRBE) and Far Infrared Absolute
Spectrophotometer (FIRAS) instrument on board the Cosmic Background
Explorer (COBE) satellite (Dwek \etal\markcite{Dwek97b}, 1997); (2)
The CFD model does not include polycyclic aromatic hydrocarbons (PAHs)
as a grain constituent. As a result it does not take the abundance of
carbon locked up in PAHs or the effect of PAHs on the UV and optical
extinction into account.  In light of these difficulties we re-examine
the theory of x-ray scattering by interstellar dust particles and the
evidence for the existence of composite fluffy dust particles in the
ISM.

We present in \S 2 a brief review of the theory, describing the
Rayleigh-Gans (RG) approximation to the x-ray scattering from a
spherical particle and examining the range of its validity.  In \S 3
we present the exact Mie solution to light scattering from a spherical
particle, and describe the differences between the Mie theory and the
Rayleigh-Gans approximation for various astrophysical situations,
including the importance of absorption of X-rays in the dust.
In \S 4 we apply the Mie theory to the observations of Nova
Cygni 1992 from Mathis \etal\markcite{Mathis} (1995), and discuss the
implications for dust models.

\section{Basic theory}

The physical situation we consider is the small-angle scattering of
X-rays emitted by a bright x-ray source from dust grains in the ISM
near the line of sight (LOS), which will create an x-ray
halo around the point source.  The scattered angles are small enough,
that for astronomical purposes, we can assume that the path traveled
by the unscattered primary X-rays is nearly identical to the paths of
scattered photons, in terms of their distances and environments.  We
discuss this assumption and its limitations in Appendix A.

The x-ray source S (see Figure~\ref{fig:scat_diag}) at distance D from
the observer O has an observed flux $F_X$ and spectrum $S(E)$\ in the
$\{ E_{min}, E_{max}\}$\ energy range, where $S(E)$\ is normalized
to unity over that interval.
\begin{figure}[tb]
\begin{center}
\includegraphics[totalheight=\bigpicsize]{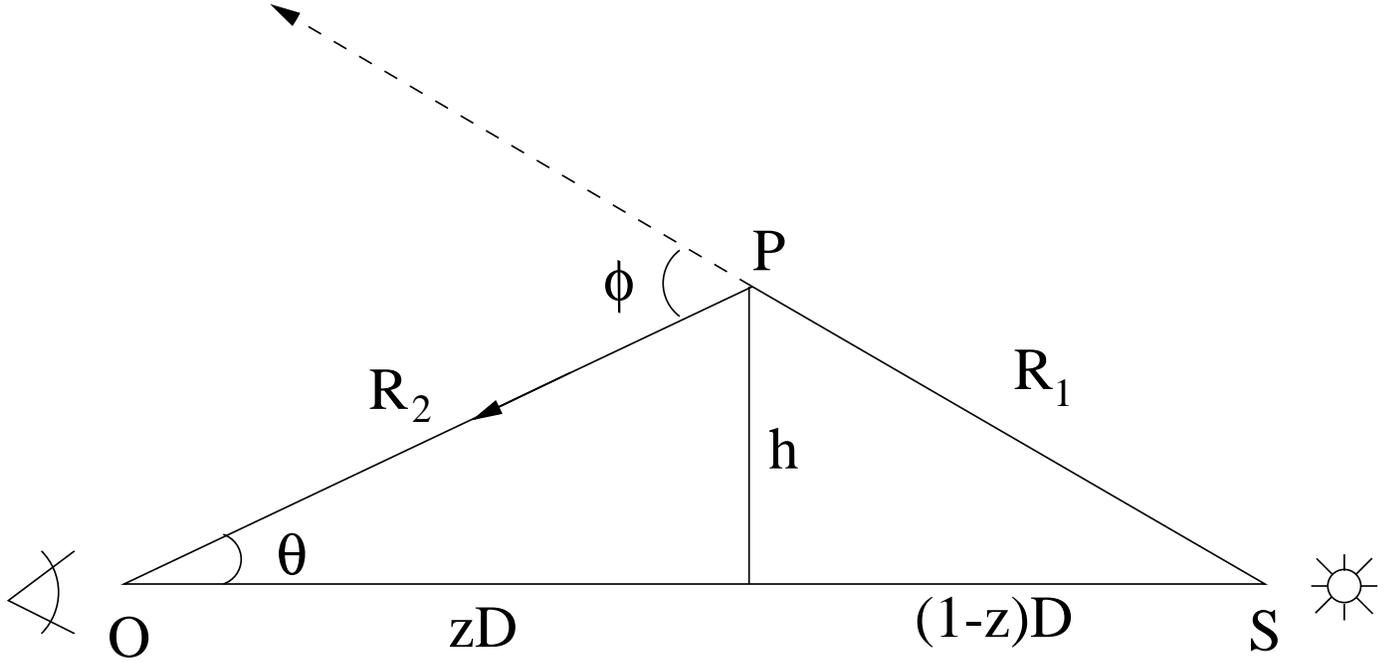}
\end{center}
\caption{Scattering diagram of an X-ray emanating from a source S 
that is scattered through an angle $\phi$\ by a dust particle
located at P towards an observer O.  The parameter $z \equiv R_2
\cos\theta / D$\ is a dimensionless quantity denoting the
projected distance to the dust particle from O along the line of
sight to S.\label{fig:scat_diag}}
\end{figure}
The emitted X-rays, traveling though the ISM scatter from a population
of dust grains with number density $n(a)\,da$\ per H atom in the
interval \{$a,a+da$\} for grain radii in the range
$\{a_{min},a_{max}\}$.  We assume that the grain size distribution is
uniform but that its total number density can vary along the LOS.  We
characterize the spatial distribution of the grains by the function
$f(z)$\ where $z \equiv R/D$\ is the normalized distance from the
observer for an actual distance $R$\ from the observer; $f(z)$\ is
normalized to unity over the range \{0,1\}  Finally, the scattered
power in a given direction is determined by the differential
scattering cross section ($d\sigma / d\Omega$), which is a function of
x-ray energy, dust grain size, and scattering angle.

In the most general form, we can write the equation for the scattered
intensity at angle $\theta$, in units of photons per cm$^{2}$
per second per steradian as (see Appendix A):
\begin{equation}
\label{scat_eq}
I_{sca}(\theta) = \int_{E_{min}}^{E_{max}} dE\,F_X\,S(E)
\int_{a_{min}}^{a_{max}} da\,\NH\,n(a) \int_{0}^{1}
{{f(z)}\over{(1-z)^2}} \Big({{d\sigma(E,a,\phi)}\over{d\Omega}}\Big)
dz .
\end{equation}
Equation (\ref{scat_eq}) apparently diverges for scattering near the
source, where $z \rightarrow 1$.  However, near the source the
contribution to the flux scattered to the observer comes from
relatively large scattering angles, for which ($d\sigma/d\Omega$) is
vanishingly small.  So in practice using an upper limit $1-\epsilon$\
instead of 1 to the position integral for some small $\epsilon$\ is
acceptable, since the integral from $1-\epsilon$\ to 1 is negligible.
Numerical experiments, with smoothly distributed dust ($f(z) \equiv
1$), a single grain size, and a single x-ray energy, show that the
integral (using the Mie solution for the scattering term) converges to
within 1 part in $10^{6}$\ when $\epsilon \equiv \theta$\ (measured in
radians).  This conclusion holds for dust sizes in the range
$0.1-1.0\,\micron$ and x-ray energies in the range $0.1-1.0$\,keV.
Using this result, we can modify equation~(\ref{scat_eq}) by changing
the upper limit on $z$\ to be $1-\theta$\ so that:
\begin{equation}
\label{scat_eq2}
I_{sca}(\theta) = F_X \NH \int_{E_{min}}^{E_{max}} dE\,S(E)
\int_{a_{min}}^{a_{max}} da\,n(a) \int_{0}^{1-\theta}
{{f(z)}\over{(1-z)^2}} \Big({{d\sigma}\over{d\Omega}}\Big) dz .
\end{equation}
which removes the numerical difficulties.

In addition, in this paper we consider only singly-scattered X-rays.
Although ML91 developed the theory of multiply-scattered x-ray halos,
the brightest x-ray halos occur when the optical depth to scattering
is low, $\tau_{sca} \sim 0.2$ (PK96).  PK96 also found that when the
optical depth for scattering was high enough that multiple scattering
was important, absorption along the LOS greatly reduced both the
source and halo intensity.

\subsection{Rayleigh-Gans Theory}

The Rayleigh-Gans approximation for the differential scattering cross
section $(d\sigma/d\Omega)$\ requires the following two assumptions.
First, that the reflection from the surface of the dust particle is
negligible, {\it i.e.}\ that $| m - 1 | \ll 1$, where $m$\ is the
complex index of refraction of the dust.  This ensures that the X-ray
enters the dust particle instead of being reflected.  The second, more
stringent requirement is that 
\begin{equation}
\label{RGlimit}
k_0 a | m - 1 | \ll 1
\end{equation}
where $k_0$\ is the wave number of the X-ray and $a$\ the radius of
the grain. This ensures that the phase of the incident wave is not
shifted inside the medium.  We will show how, in this limit and for
sufficiently small scattering angles, it is possible for the waves
scattered throughout the dust grain to add coherently.  In this
situation, the intensity of the scattered waves is proportional to the
number of scattering sites squared; {\it i.e.}:
\begin{equation}
\label{SimpleProp}
I \propto N^2 \propto \rho^2 a^6
\end{equation}
where $\rho$\ is the mass density of the dust grain, and $a$\ its
radius.

A useful expression for the RG approximation assumptions can be derived
using the ``Drude approximation'' to $|m-1|$\ for a collection of free
electrons responding to an oscillating external field.  The Drude
approximation (Bohren \& Huffman\markcite{BohrenHuffman}, 1983,
p. 253) is
\begin{equation}
\label{Drude}
m \approx \sqrt{1 - {{\omega_p^2}\over{\omega^2 + i\gamma\omega}}}
\end{equation}
where $\omega$ is the circular frequency of the wave, $\omega_p^2 = 4
\pi c^2 n_e r_e$\ is the plasma frequency of the medium (for electron
number density $n_e$, and where $r_e$\ is the classical radius of the
electron). The $\gamma$\ term is a damping term due to electronic
interactions with impurities or lattice vibrations.  Assuming $|m - 1|
\ll 1$, we can rewrite (\ref{Drude}) to get
\begin{equation}
| m - 1 | \approx {{n_e r_e \lambda^2}\over{2\pi}} \Big({{1}\over{1 +
{{i\gamma}\over{\omega}}}} \Big)
\label{DrudeApprox}
\end{equation}
where $n_e$\ is the density of electrons, $r_e$\ is the classical
radius of the electron, and $\lambda$\ the wavelength of the photon.
Combining (\ref{DrudeApprox}) with (\ref{RGlimit}), and ignoring
damping ({\it i.e.}, $\gamma = 0$) we get:
\begin{equation}
\label{BasicLimit}
a n_e r_e \lambda \ll 1.
\end{equation}
Equation~(\ref{BasicLimit}) can be expressed in terms of the dust mass
density $\rho$\ by assuming that the atomic constituents of the dust
have one electron per 2 nucleons, so $n_e = \rho/(2m_H)$.  The
coherency requirement for the validity of the RG approximation then becomes:
\begin{equation}
{{a_{\micron}\over{E_{keV}}}} \Big( {{\rho}\over{3\,\hbox{g\,cm}^{-3}}}
\Big) \ll 0.316.
\label{RuleOfThumb}
\end{equation}
A ``rule of thumb'' for normal interstellar dust grains is that the RG
approximation is valid only if the energy of the x-ray in keV is
significantly larger than the grain radius in $\mu$m.

Within this limit, the RG approximation completely describes the
scattering of photons in a simple analytic form.  The exact RG
approximation for the differential scattering cross section is given
in MG86, and shown below.  The exact result is in terms of the
spherical Bessel function $j_1$, but can be well-approximated by a
Gaussian function:
\begin{eqnarray}
\Big({{d\sigma}\over{d\Omega}}\Big)(E, a, \phi) & = &2a^2\Big({{2\pi
a}\over{\lambda}} \Big)^4| m - 1|^2 \Big({{j_1(y)}\over{y}} \Big)^2 (1
+ \cos^2 \phi) \label{RGbessel} \\ 
& \approx & 2a^2\Big({{2\pi a}\over{\lambda}} \Big)^4| m - 1|^2 {1\over9}
\exp\Big(-{{y^2}\over{2\sigma^2}} \Big) (1 + \cos^2 \phi)
\end{eqnarray}
where $y = (4\pi a/\lambda) \sin(\phi/2)$\ and $\sigma$\ is a
function of the energy and size of the dust particle.  We see here the
$a^6$\ dependence described earlier; the $\rho^2$\ dependence is
implicit in the $|m-1|^2$\ term (see eq. \ref{DrudeApprox}).

\begin{figure}[tb]
\includegraphics[totalheight=3in]{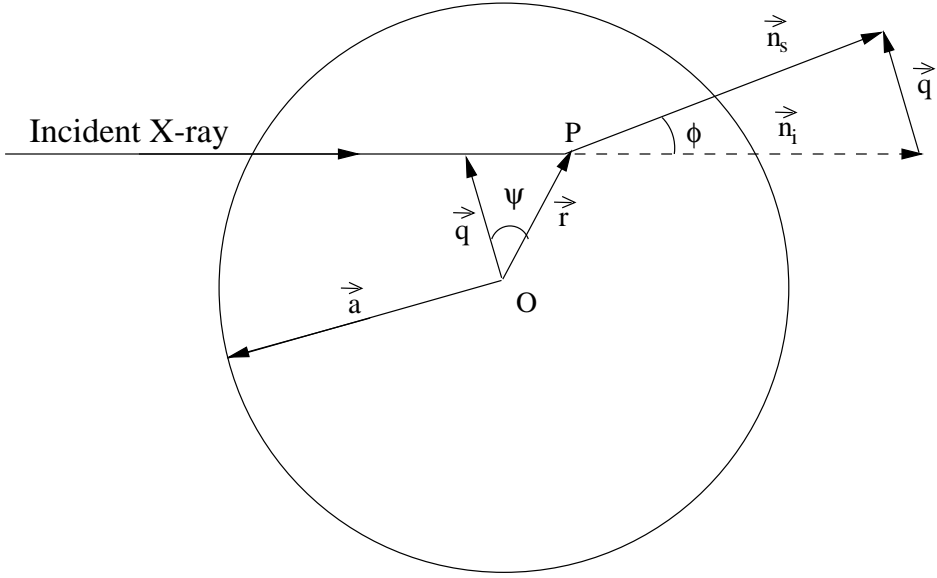}
\caption{An X-ray traveling in direction $\vec{n_i}$\ impinging upon a
dust particle of radius $a$, interacting at a point P located at a
distance $r$\ from the center of the particle O, and scattering in the
direction $\vec{n_s}$.\label{DustScatDiag}}
\end{figure}

Insight into the mechanism and the maximum angle for coherent
scattering can be derived from the fundamental physics of the
scattering.  Photons scatter from charged particles, and the
differential scattering cross section for photons interacting with a
collection of charged free particles is given by
Jackson\markcite{Jackson} (1975) (eq. 14.111):
\begin{equation}
\label{JacksonEq}
{{d\sigma}\over{d\Omega}} = | \sum_j {{e_j^2}\over{m_j c^2}}
e^{i \vec{q} \cdot \vec{r_j}} |^2 | \vec{\epsilon}^* \cdot
\vec{\epsilon_o} |
\end{equation}
where the sum is taken over all the charges in the medium.  Here $e_j$
is the charge of the $j$th particle, $m_j$ is its mass, and $r_j$ is
its position relative to the center of the dust particle (see
Figure~\ref{DustScatDiag}).  We define $\vec{q} \equiv (\omega
\hat{n_s} - \omega_0 \hat{n_i})/c$, where $\hat{n_i}$ and $\hat{n_s}$
are unit vectors in the initial and scattered directions respectively,
$\omega_0$ and $\omega$ are the circular frequencies of the X-ray
outside and inside the medium respectively, and $c$\ is the speed of
light.  The initial and scattered polarization is given by
$\vec{\epsilon_0}$ and $\vec{\epsilon^*}$ respectively; however, we
will not consider x-ray polarization in this paper and will therefore
ignore these terms.

We see immediately that all the scattering is by electrons; scattering
by protons is strongly suppressed because of their large mass.
Figure~\ref{DustScatDiag}\ diagrams the situation for one particular
scattering site in a spherical dust grain.

Equation (\ref{JacksonEq}) is strictly valid only for free electrons.
However, X-rays are substantially more energetic than the electronic
energy bands in the solids so the electrons in the dust grain can
be treated as free (Jackson, 1975, p. 683).  Absorption, however, is
not entirely negligible even above 1 keV as there are a number of
elements with K-shell electrons that can absorb at these energies (see
\S 3.2).

We can now show how to derive the limiting angle for coherent
scattering.  If $|\vec{q} \cdot \vec{r_j}| \ll 1$\ in
Equation~(\ref{JacksonEq}), the exponent term reduces to unity.  Then
the sum is simply $N^2 e^2 / (m_e c^2)$, where $N$\ is the total
number of electrons in the dust grain.  Physically, this result is due
to each electron scattering independently and the resulting waves
adding coherently.  The intensity therefore depends upon the {\it
number}\ of electrons in the scattering medium, and not on the {\it
spacing}\ or {\it density}\ of the electrons.

Expanding terms in $|\vec{q} \cdot \vec{x_j}| \ll 1$, we get, using
the notation of Figure~\ref{DustScatDiag}:
\begin{eqnarray}
\label{OmegaEq}
\nonumber |\vec{q} \cdot \vec{r}| & = & |({{\omega}\over{c}} \hat{n_s} -
{{\omega_0}\over{c}} \hat{n_i}) \cdot \vec{r}| \\
\nonumber & = & | ({{\omega}\over{c}} \hat{n_s} - {{\omega_0}\over{c}}
\hat{n_i}) |\cdot | \vec{r} |\cdot |\cos\psi| \\
 & = & \Big[ ({{\omega}\over{c}})^2 + ({{\omega_0}\over{c}})^2 -
2{{\omega \cdot \omega_0}\over{c^2}} \hat{n_s}\cdot\hat{n_i}
\Big]^{1/2} \cdot r \cdot | \cos\psi |.
\end{eqnarray}
The complex optical constant for the dust grain relates $\omega$\ and
$\omega_0$ via $\omega = m\omega_0$.  Substituting in (\ref{OmegaEq})
we get
\begin{equation}
|\vec{q} \cdot \vec{r}|  = k_0 \Big[ m^2 + 1 - 2 m \cos\phi \Big]^{1/2}
\cdot r \cdot |\cos \psi| 
\end{equation}
where $k_0 = \omega_0 / c$.  The position vector $\vec{r}$\ takes on
all values between $0 < r < a$, where $a$\ is the radius of the dust,
and $0 < |\cos\psi| < 1$.  Replacing $r |\cos \psi|$\ with its maximum
value, $a$, the requirement that $|\vec{q}\cdot\vec{r}| \ll 1$\
becomes
\begin{equation}
\label{FirstRewrite}
| (m^2 + 1 - 2 m \cos\phi) |^{1/2} k_0 a \ll 1.
\end{equation}
Note that Equation~(\ref{FirstRewrite}) is equivalent to
Equation~(\ref{RGlimit}) for $\phi \rightarrow 0$, so if the RG
assumptions are met coherent scattering will occur for at least some
scattering angles.  We can rewrite Equation~(\ref{FirstRewrite}) using
some trigonometry and the already-stated assumption that $|m - 1| \ll
1$\ to get a better limit on the scattering angle: 
\begin{equation}
2 k_0 a \sin({{\phi}\over2}) \ll 1.
\end{equation}
In the x-ray regime $k_0 \sim 5000\,E_{keV} \micron^{-1}$.  Thus $k_0
a$\ will much larger than unity for normal interstellar dust grains
and $\sim$ 1 keV X-rays, so we can use the small angle approximation
for $\sin(\phi/2)$.  Waves scattered at angles where $\phi \ll 1 / k_0
a$, when the RG assumptions are met, will have intensities
proportional to $N^2$, the number of electrons in the grain squared.

To summarize, the RG approximation requires that the dust neither
reflects photons nor significantly affects their phases.  The induced
scattering wave generated at the entrance to the grain must add in
phase with one created at the exit.  If it does not, then coherency is
lost, and the amplitude of the scattered wave never becomes
proportional to the number of electrons squared in the entire dust volume
squared, but to a fraction of that number.

Furthermore, the combination of the RG approximation with the Drude
approximation assumes that absorption is negligible at all energies.
This is approximately valid when $E > 1$\,keV, although there are
astronomically abundant elements in dust with atomic energy levels
above 1 keV are magnesium (K edge, 1.294 keV), aluminum (K edge, 1.550
keV), silicon (K edge, 1.828 keV), and iron (K edge, 7.083 keV).
Below 1 keV, however, there are many K and L shell edges which can
absorb X-rays, an effect which will be considered in \S 3.2.  The
breakdown of the RG approximation and the effect of absorption require
the use of full Mie theory to correctly calculate the intensity of the
scattered light.

\section{Mie Theory}

Mie theory calculates the exact solution for the scattering and
absorption of light from a spherical particle (van de
Hulst\markcite{Hulst}, 1957).  Unlike the Rayleigh-Gans approximation,
it does not assume all the scattering sites will add coherently and as
a result, Mie theory can correctly calculate the scattered intensity
for particles even when $a_{\mu m} \gtrsim E_{keV}$.  In the limit where
$ k_0 a | m - 1 | \ll 1$\ the Mie solution recovers the result of the
Rayleigh-Gans approximation.  Another advantage is that absorption can
be included easily, using measured values for the index of refraction.
However, although the Mie solution is exact for light scattering from
a sphere, unlike the RG approximation it is not expressible in a
useful analytic form.

The Mie solution requires careful numerical computation, especially
for large values of the ``size parameter'' $x \equiv 2\pi a/ \lambda
\approx 5000\,a_{\micron}\,E_{keV}$ where the series solution requires
summation over a large number of terms to converge.  We use the Mie
code developed by Wiscombe\markcite{Wiscombe} (1979, 1980), which was
written for atmospheric use but has been tested for size parameters up
to 20,000, sufficient for our problem.  The observed scattered
intensity is obtained by integrating the intensity of the light
scattered by a single dust particle to the observer, over the grain
size and spatial distribution along the LOS (see eq. (\ref{Gen_Scat}) in
the Appendix).  These calculations can be greatly facilitated by
pre-calculating tables of ($d\sigma( E, a, \phi )/d\Omega $) for a
range of energies, dust sizes, and scattering angles. \footnote{The
code needed to generate such tables, and related utilities, may be be
obtained from the authors.}

We must point out, however, that even the Mie solution is only of an
approximate nature, since dust particles are neither spherical nor
compositionally homogeneous.  We know that the first assumption is
incorrect, since only non-spherical grains can polarize light, as
grains have been observed to do ({\it e.g.}\ Mathis, 1990).  The
second assumption is particularly important, since the wavelength of a
1 keV photon is only 1.24\,\AA, substantially smaller than the radius
of any dust grain.  Thus, the X-ray's path through the grain will
determine which dust grain constituents it interacts with, and so the
x-ray scattering will be affected by the grain's exact morphology.
Exploring this possibility in detail would require a discrete dipole
approximation to the dust grain constituents ({\it e.g.}\ Draine,
1988), which is beyond the scope of this paper.

\subsection{Comparison to the Rayleigh-Gans Approximation}

\begin{figure}[tb]
\begin{center}
\includegraphics[totalheight=\smpicsize]{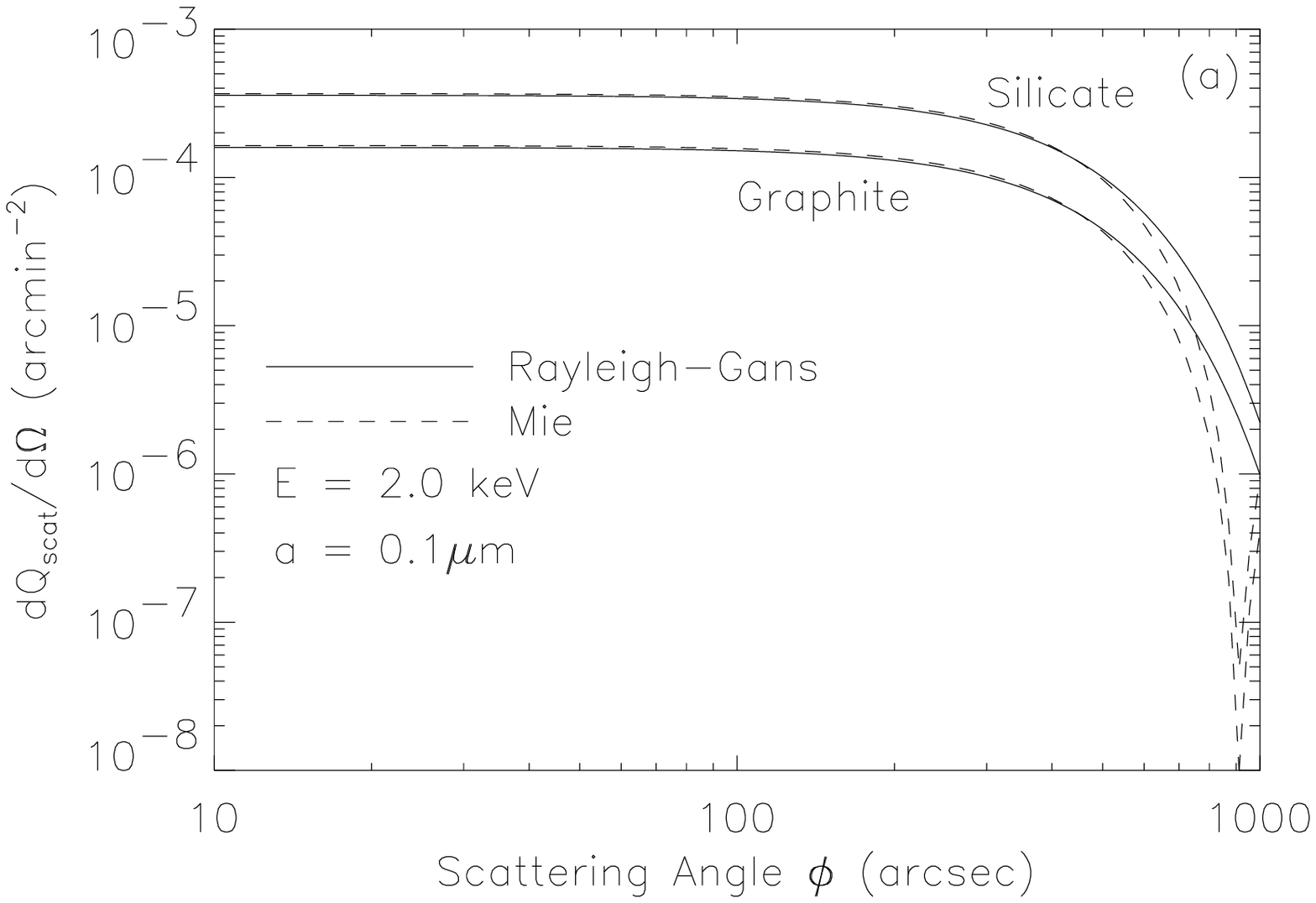}
\includegraphics[totalheight=\smpicsize]{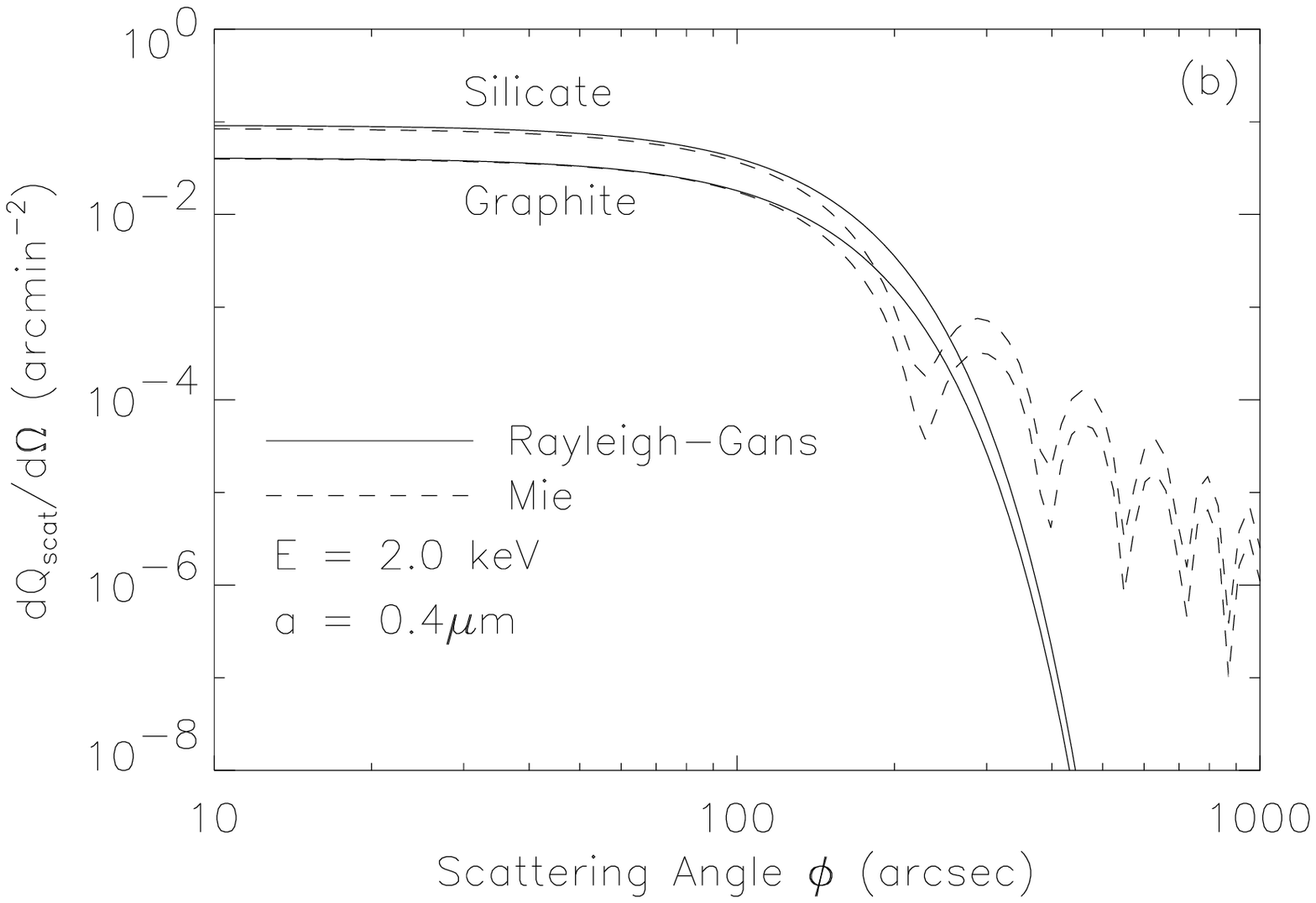}
\end{center}
\caption{(a) The differential scattering cross sections
$dQ_{sca}/d\Omega = (d\sigma/d\Omega)/\pi\,a^2$, for a 2 keV X-ray
scattering from a $0.1\,\mu$m dust grain, for both graphite and
silicate dust grains.  The solid lines show the Rayleigh-Gans
approximation using the Gaussian approximation; the dashed lines show
the equivalent Mie calculation.  In this energy and $\phi$-range the
two methods should give identical results; however, use of the
Gaussian approximation leads to deviations at large angles.  (b) Same,
for $0.4\,\mu$m dust grains.\label{fig:Q_02}}
\end{figure}

The observed halo intensity can be calculated for any x-ray energy and
dust grain size, using either RG approximation or the Mie solution.
Any variation in the result is due solely to the differential
scattering cross section $(d\sigma/d\Omega)$\ term.  In
Figure~\ref{fig:Q_02} we plot $(dQ_{scat}/d\Omega) \equiv
(d\sigma/d\Omega)/\pi a^2$\ for 2 keV X-rays scattering from a
single dust grain, as a function of the scattering angle $\phi$.  The
differential scattering cross section was divided by the geometrical
cross section to show the relative effect of grain size on the
scattering intensity.  Figure~\ref{fig:Q_02}(a) presents the results
for both a graphite or silicate dust grain of radius $0.1\,\mu$m and
Figure~\ref{fig:Q_02}(b) is similar but for grains of size
$0.4\,\mu$m.  The mass densities of the two types of dust, graphite
and silicate, were 2.2\,g\,cm$^{-3}$ and 3.3\,g\,cm$^{-3}$
respectively.  The optical constants were calculated using the Drude
approximation, ignoring absorption, both for the RG and the Mie cases,
so that the variation in the figures is due solely to the approximate
nature of the RG solution.

Figure~\ref{fig:Q_02} shows that at 2 keV, the Gaussian fit to the
Rayleigh-Gans approximation is adequate to the task; it deviates
substantially from the Mie solution only at large angles, after the
scattering intensity has dropped by at least an order of magnitude.
The deviation itself is largely due to the Gaussian approximation. 
Using the RG approximation with the spherical Bessel function
duplicates the ripples in the large-angle scattering of the
$0.4\,\mu$m grains and the faster drop seen for the 0.1\,$\mu$m
grains.

\begin{figure}[tb]
\begin{center}
\includegraphics[totalheight=\smpicsize]{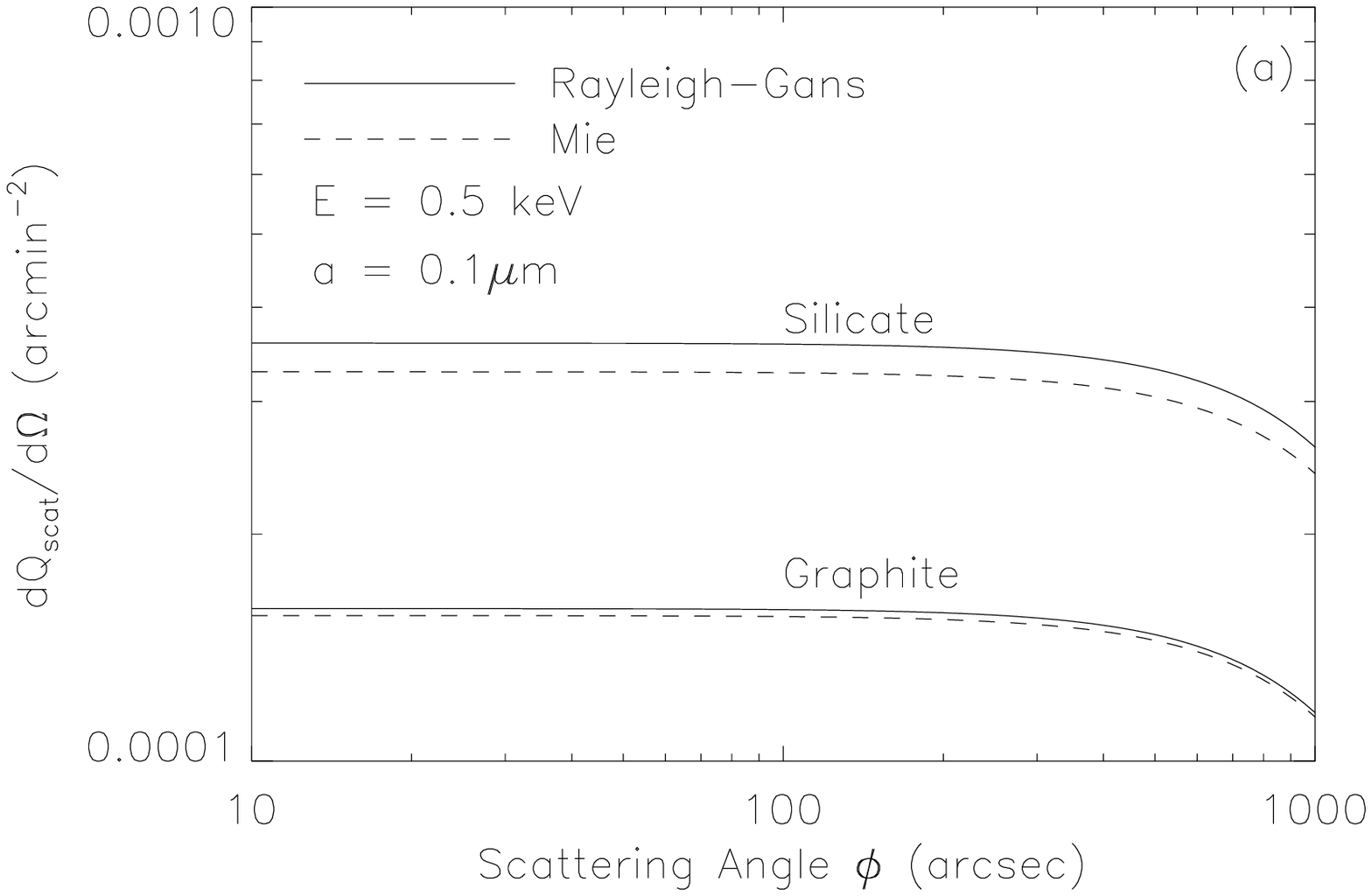}
\includegraphics[totalheight=\smpicsize]{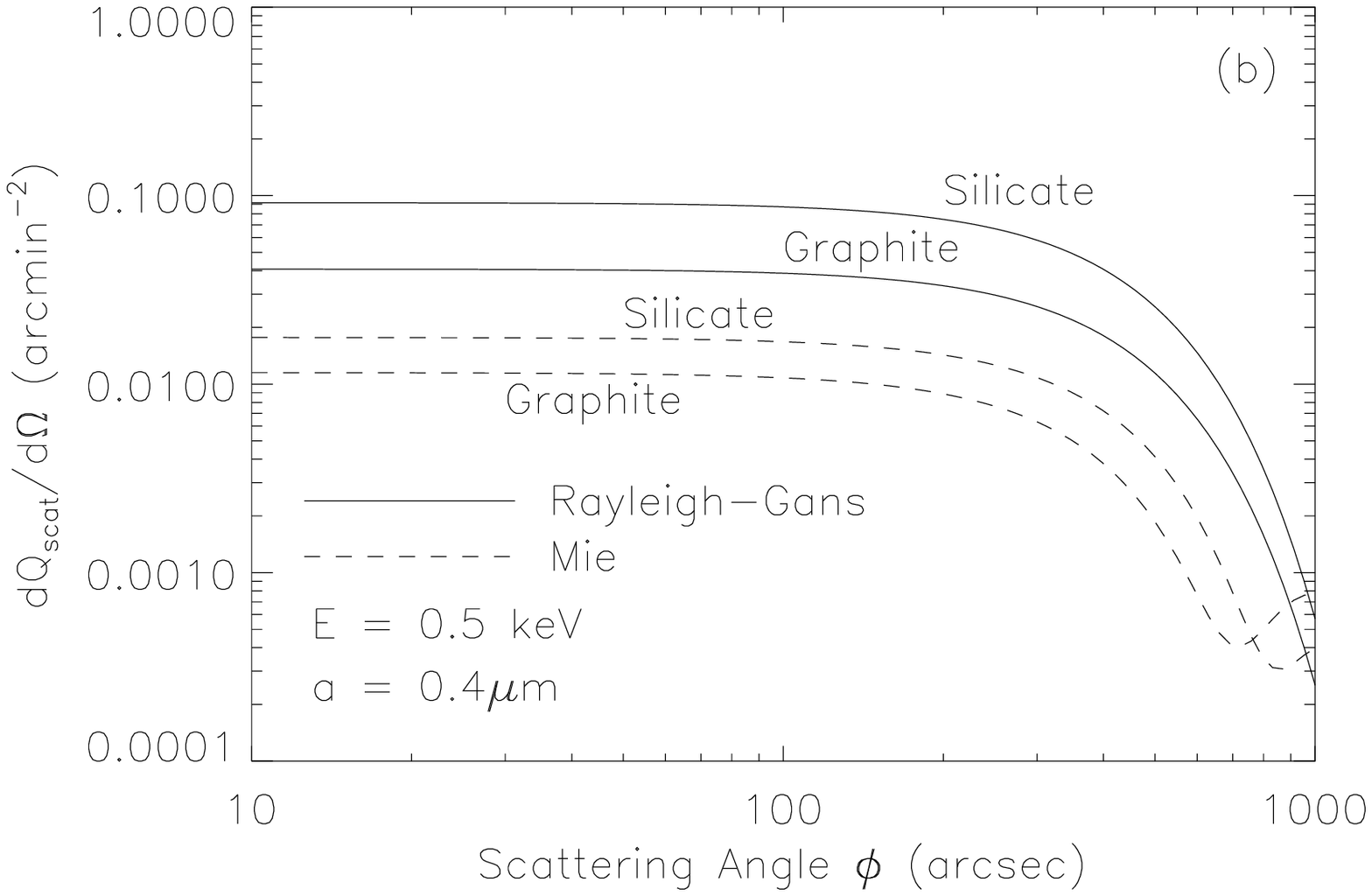}
\end{center}
\caption{(a) Same as Figure \protect{\ref{fig:Q_02}}, for 0.5 keV
X-rays.  The RG approximation is an acceptable fit to the Mie
calculations. (b) The same as (a), for a $0.4\,\mu$m dust grain.  The
assumptions necessary for the RG approximation do not hold, and the
results show that the differential scattering cross sections
calculated using the Mie and RG approximation disagree substantially
at all scattering angles.\label{fig:Q_005}}
\end{figure}

When the requirement that $k_0 a | m - 1 | \ll 1$ is not satified, the
agreement between the RG and the Mie solution becomes substantially
worse.  Figure~\ref{fig:Q_005} is identical to Figure~\ref{fig:Q_02},
but for X-rays of energy 0.5 keV.  The ``rule of thumb'' suggests that
in this case the RG approximation will hold for $0.1\,\mu$m radius
dust, but fail for the $0.4\,\mu$m radius dust. Figure~\ref{fig:Q_005}
confirms this, showing an acceptable match between the RG and the Mie
calculations for the smaller grains, but showing far more total
scattering with the RG approximation than in the Mie solution for the
larger grain size.  In this case, using the exact Bessel function
solution does not help.  The larger dust grain has more scattering
sites, and the RG approximation assumes all the scattered waves all
add coherently; the Mie solution shows that for large dust sizes this
coherency assumption fails, and the scattered intensity therefore
drops.  Figure~\ref{fig:Q_005} suggests that the approximation fails
abruptly, as might be predicted from the exponential nature of
equation~(\ref{JacksonEq}).

\subsection{The Effect of Absorption}

We have so far used the Drude approximation to the optical constants,
ignoring any damping, in both the RG approximation and the Mie
solution.  Although the Drude approximation works well above 1 keV,
real dust grain constituents have many K and L shell electrons with
energy levels that are in the range $0.1 - 1$\,keV, and these
electrons will absorb soft X-rays.   The effect of absorption can be
included by using the atomic scattering factor $F(E)$\ (Henke
\markcite{Henke}, 1981) as a modifier to the Drude approximation:
\begin{equation}
| m  - 1 | = {{n_e r_e \lambda^2}\over{2\pi}} | F(E) / Z |
\end{equation}
where $Z$\ is the atomic number of the element.  However, we simply
use the measured optical constants which include the absorption
effects.  For silicate particles (\{MgFeSi\}O$_4$) we used the optical
constants of Martin \& Rouleau\markcite{MartinRouleau} (1991), and for
graphite particles we used Rouleau \& Martin\markcite{RouleauMartin}
(1991).  These values are a smooth extension of the Draine \&
Lee\markcite{DraineLee} (1984) and Laor \& Draine\markcite{LaorDraine}
(1993) results into the FUV and x-ray regime.  At x-ray energies,
there is no distinction between graphite and amorphous carbon, so this
complication need not be considered.  We also considered composite
dust particles as described in Mathis (1996); for this case, we used
the Bruggeman rule (Bohren \& Huffman, 1983) to calculate the optical
constants of the mixture.

\begin{figure}[tb]
\begin{center}
\includegraphics[totalheight=\smpicsize]{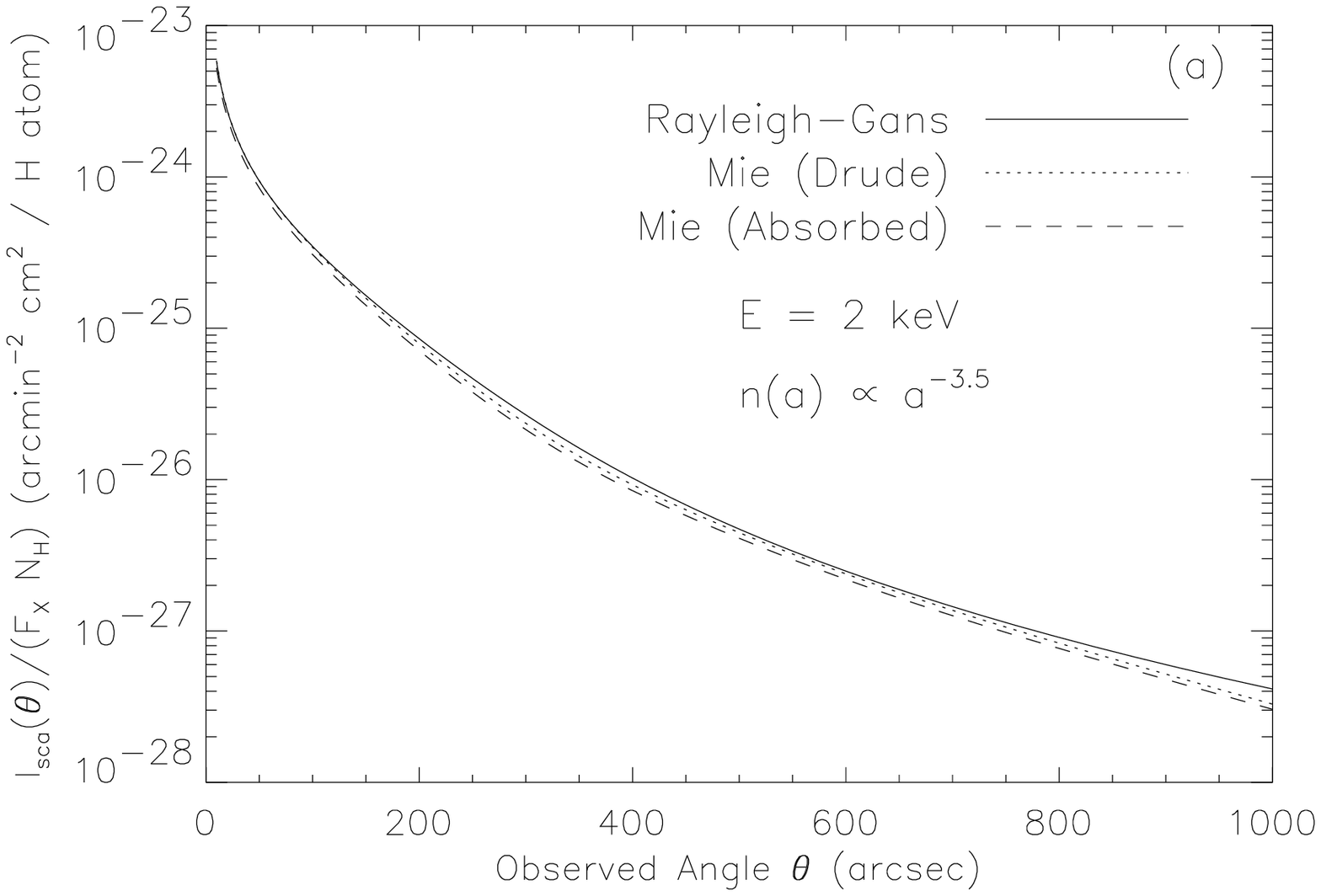}
\includegraphics[totalheight=\smpicsize]{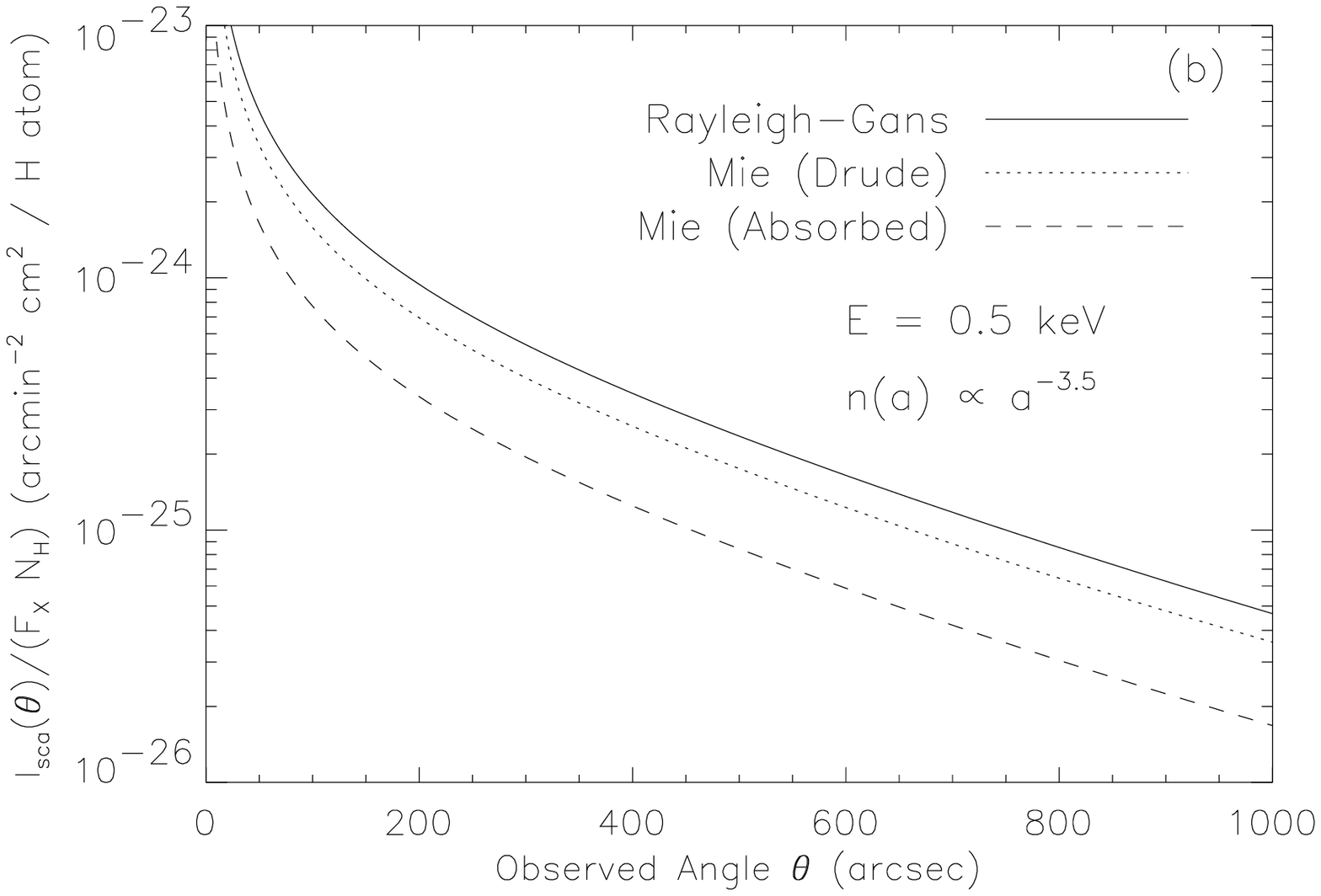}
\end{center}
\caption{The observed halo intensity from a 2 keV x-ray source
scattering from evenly-distributed silicate and graphite dust grains
with an $n(a) \propto a^{-3.5}$\ size distribution.  The results are
shown using the RG approximation, the Mie solution with the Drude
approximation to the index of refraction ({\it i.e.}, without
absorption) and for the Mie solution using measured optical
constants.\label{fig:Abs}}
\end{figure}

Figure~\ref{fig:Abs}(a) plots the observed halo intensity for 2 keV
X-rays and a distribution of grain sizes, calculated three different
ways.  We presented the observed halo, rather than showing
$(d\sigma/d\Omega)$\ for a single dust grain size as was shown in
Figures~\ref{fig:Q_02} and~\ref{fig:Q_005}, to show the magnitude of
the observable differences between the methods.  For 2 keV X-rays,
these differences are clearly small.  The halo intensity, given by
equation~(\ref{scat_eq}), was found by integrating
$(d\sigma/d\Omega)$\ over an evenly distributed populates of dust
particles ($f(z)\equiv1$) characterized by a size distribution $n(a)
\propto a^{-3.5}$ with $\{a_{min},a_{max}\} = \{0.005\mu m,0.25\mu
m\}$.  The grains themselves consisted of a mix of silicate and
graphite grains with dust to hydrogen mass ratios of 0.0057 and 0.0032
respectively.  

We note that the choice of the upper limit to the size distribution,
$0.25\mu$m, significantly affects the result.  With this power law
size distribution, most of the mass is in the largest grains.  In
addition, the distribution of the largest dust particles are not
well-determined by extinction measurements, since when the grain size
is much larger than the wavelength the extinction becomes independent
of wavelength.  X-ray halos are therefore a particularly useful and
sensitive probe of the number and size of large dust grains.  In the
RG approximation the intensity is proportional to the sixth power of
the radius of the dust.  The Mie solution also predicts larger dust
grains scatter more than small grains, although not by as large a
power (see Figures~\ref{fig:Q_005}(a) and (b)).  In general, however,
the source function of the scattered intensity is heavily weighted
towards large dust grains.

Figure~\ref{fig:Abs} shows the cumulative effects of the
approximations used in calculating the observed halo intensity.  The
simple RG approximation (with the Gaussian fit to the Bessel function
and the Drude approximation to the optical constants) predicts the
largest intensity.  The exact Mie solution, using the Drude
approximation, reduces the intensity since not all the scattered waves
add coherently.  The Mie solution, using actual optical constants
which include absorption, has the lowest intensity as some of the
impinging X-rays are absorbed in the dust.

The effect of absorption can be much larger, as shown in
Figure~\ref{fig:Abs}(b), which is identical to Figure~\ref{fig:Abs}(a)
but for 0.5 keV X-rays.  The difference between the RG and the Mie
solution for a realistic dust grain size distribution is quite
noticeable, and Figure~\ref{fig:Abs}(b) shows that absorption is very
important at this energy.  We note in passing that not all of the
absorbed x-ray's energy will be deposited in the dust (Dwek \&
Smith\markcite{DwekSmith}, 1996) since the ejected photo-electron may
deposit only a fraction of its energy in the solid.

We have shown that the accuracy of the RG approximation is size
dependent and is in general less accurate at larger grain sizes.  The
interstellar grain population is characterized by a mixtures of grain sizes and
it is therefore interesting to apply Mie theory to an astronomical
situation.  

\subsection{The Energy Dependence}

\begin{figure}[tb]
\begin{center}
\includegraphics[totalheight=\bigpicsize]{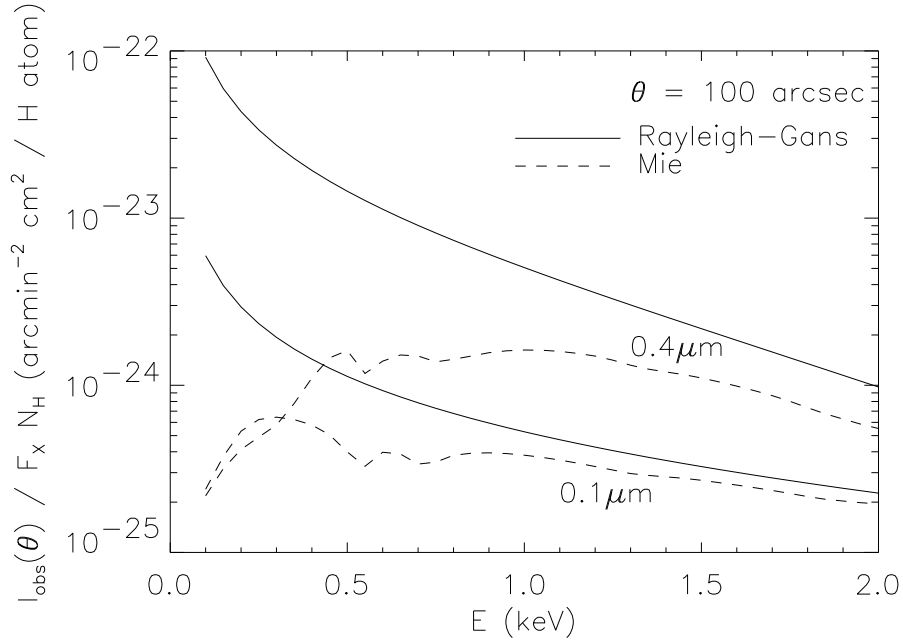}
\end{center}
\caption{The scattered intensity at 100$''$\ for silicate grains of
radius 0.1 and 0.4 $\mu$m, calculated both by the RG approximation
(without absorption) and the Mie solution (including absorption) as a
function of x-ray energy.  For $E \gtrsim 1$\,keV, the two
methods show the same energy dependence; below it, absorption modifies
the $E$\ dependence, an effect that becomes more pronounced at 0.4\,$\mu$m.
\label{fig:Mie_RG_comp} }
\end{figure}

In Figure~\ref{fig:Mie_RG_comp}, both the Mie and RG methods were used
to calculate the intensity of scattered X-rays from silicate grains as
a function of x-ray energy, at the observed scattering angle
$\theta = 100''$.  We see that for higher energies, the shape of
the curves for the two methods begins to agree when $E \gtrsim
1$\,keV, but below this energy there are substantial differences.  The
two methods never agree exactly because of absorption, which is more
important for larger grains but is not negligible even for the 0.1
$\mu$m grain.  The presence of absorption lines in the grains can also
be seen in the halo shape, which has irregular peaks and valleys in
the Mie case.

We conclude that in many astronomical situations using the RG
approximation to calculate the halo intensity systematically
overestimates the amount of x-ray scattering from a given dust size
distribution.  Consequently, the RG approximation cannot be used for
x-ray energies below 1 keV, except as an upper limit to the halo
intensity.  Even for energies above 1 keV, RG is only an upper limit
since absorption is never totally negligible, and will reduce the
total intensity of scattered light.

\section{Application and Discussion}

The total scattered light intensity will in practice depend upon the
spectrum of the source $S(E)$, the spatial distribution of dust grains
$f(z)$, and the grain size distribution $n(a)$.  We can measure the
x-ray spectrum $S(E)$, directly, and we will assume that dust grains
are evenly distributed, so that $f(z) \equiv 1$.  The last parameter,
the dust grain size distribution $n(a)$, is not well known, especially
for larger size grains.  Currently, it is measured by fitting the
extinction of visible and UV starlight.  We will present results using
two different grain models: (1) a bare core silicate and graphite mix,
with an $a^{-3.5}$\ power law size distribution (following Mathis,
Rumpl, \& Nordsieck\markcite{MRN}, 1977)(MRN); and (2) the recent
composite fluffy dust model of Mathis\markcite{Mathis96} (1996)(M96) a
distribution centered near 0.1 \micron, consisting of composite
silicate and graphite.  We now apply the RG and Mie solutions with
these grain models to the observed scattering from the x-ray source
(Nova Cygni 1992).

\subsection{Nova Cygni 1992}

Nova Cygni 1992 exhibited a strong x-ray halo in the ROSAT PSPC (Mathis
\etal\markcite{Mathis}, 1995).  The observed spectrum was formally fit
with a blackbody with temperature $E = kT \sim 22$\,eV, absorbed by an
equivalent hydrogen column density of $\NH =
4.25\times10^{21}$\,cm$^{-2}$.  Mathis \etal\markcite{Mathis} (1995)
do not claim this is a true model of the nova, only that it is simple
to calculate and fits the observed spectrum reasonably well.  The observed
spectrum peaked at about 0.5 keV, and with this choice of assumed
spectrum most of the photons from the source were absorbed.  However,
this column density is not the best choice for use in the scattering
intensity equation~(\ref{scat_eq}), since the value of \NH\ is highly
dependent upon the underlying true spectrum.  In addition, the
hydrogen column density measured from x-ray absorption depends upon
the helium and metals along the LOS, and is not directly related to
the number of dust grains along the LOS.  Therefore, x-ray absorption
measures of the hydrogen column density may not accurately reflect
either the actual hydrogen column density or the dust grain column
density.

The scattered halo intensity scales linearly with the column density
of dust grains, which can be found by measuring the extinction along
the LOS.  Mathis \etal\markcite{Mathis} (1995) (and references
therein) estimated the reddening E(B-V) to be between 0.19-0.31 mag.
Using the standard extinction law value of $R=3.1$, this translated
into a hydrogen column density in the range $\NH = 1.1-1.8
\times10^{21}$\,cm$^{-2}$.  There is a significant difference between
the x-ray and extinction derived hydrogen column densities.  The
former depends upon the source spectrum over a broad range of
energies, whereas the latter depends on the source spectrum over a
much narrower range (B and V bands), and provides a more direct
measurement of the dust column density along the LOS.  We therefore
prefer the extinction determined hydrogen column density, and use a
value of $\NH = 1.45\times10^{21}$\,cm$^{-2}$.  For the Nova Cygni
source Mathis \etal\markcite{Mathis} (1995) noted that this column
density creates only a small optical depth for scattering, so the
assumption of single-scattering is adequate.  Nonetheless, the halo
was clearly visible in the data to angles greater than $800''$\ before
it merged with the background.

Mathis \etal\markcite{Mathis} (1995) fitted the halo using the RG
approximation for the differential scattering cross section and found
that an MRN model with a column density chosen to produce the observed
visual extinction would scatter more X-rays than were seen.  Noting
that the intensity of the scattered light is proportional to the grain
density squared, they proposed a model consisting of composite fluffy
dust (CFD) particles with substantial amounts of vacuum which would
scatter fewer X-rays but would nevertheless produce the same observed
visual extinction.  The amount of light scattered from a fluffy grain
is less than that from the same size dense grain, since the fluffy
grain has fewer electrons with which to scatter the X-rays.  This
reduction in scattering remains even when the total amount of mass in
dust is constant, since the intensity is proportional to the number of
electrons in each grain squared.  Therefore, having twice as many
grains which are 50\% vacuum will lead to a net 50\% drop in the
intensity--when the RG approximation is valid.
  
In the 0.5 keV energy range, however, the RG approximation is not
valid for dust grains larger than $\sim 0.1 \mu$m, for two reasons.
First, the scattered waves will not add coherently as the dust grain
material will affect their phases substantially, and secondly, absorption
in the dust grain will remove some of the X-rays altogether.  As a
result, using the RG approximation with the no-damping Drude
approximation to the optical constants will tend to substantially
overpredict the amount of scattered light and so the Mathis
\etal\markcite{Mathis} (1995) result must be re-evaluated using the
Mie solution.

To examine the robustness of the conclusions derived by Mathis
\etal\markcite{Mathis} (1995) for the Nova Cygni data, we modeled the
same data using Mie theory for the standard MRN model, and the
newly-proposed CFD model (M96).  The parameters of the two models are
presented in Table~\ref{DustModels}.  The two dust models produce an
overall equally good fit to the average interstellar extinction curve
(Dwek\markcite{Dwek97}, 1997).  However, the CFD model produces a
visual extinction that is 6.7\% higher than the MRN model, so we
normalized the x-ray scattering intensity downward by the same amount.
With this renormalization the two models produce the same amount of
visual extinction towards the x-ray source.  We limited our results to
the case of evenly distributed dust, {\it i.e.}\ $f(z) \equiv 1$.

\begin{table}[tb]
\caption{Dust grain parameters\label{DustModels}}
\begin{tabular}{llcllll} \hline \hline
Model &Component &$n(a)$ & $\rho$ & C/H & Si/H &f$_{\rm vacuum}$ \\ &
& & (g cm$^{-3}$) & (ppM) & (ppM) & \\ \hline
MRN$^1$  & Graphite  & $a^{-3.5}$ & 2.2  & 270 & 0  & 0 \\
     & Silicate  & $a^{-3.5}$ & 3.3  & 0   & 33 & 0\\
CFD$^2$  & Graphite  & $\ll 0.03\,\micron$ & 2.25 & 55  & 0  & 0\\
     & Silicate  & $\ll 0.02\,\micron$ & 3.3  & 6.5 & 0  & 0 \\
     & Composite$^3$ & $a^{-3.5} e^{-(0.0033a + 0.437/a + 50 a^2)}$ 
                              & 1.55 & 105 & 26 & 0.45 \\ \hline
\end{tabular}

$^1$ From Mathis, Rumpl, and Nordsieck\markcite{MRN} (1977), with
an extended size range $\{a_{min},a_{max}\}~=~\{0.005\micron,0.25\micron\}$ \\ 
$^2$ Composite fluffy dust model (CFD), based on
Mathis\markcite{Mathis96} (1996) and Dwek\markcite{Dwek97} (1997) \\  
$^3$ CFD model contains amorphous carbon and silicates; for the
size distribution $n(a)$, $a$\ is measured in microns.
\end{table}

\begin{figure}[tb]
\includegraphics[totalheight=\bigpicsize]{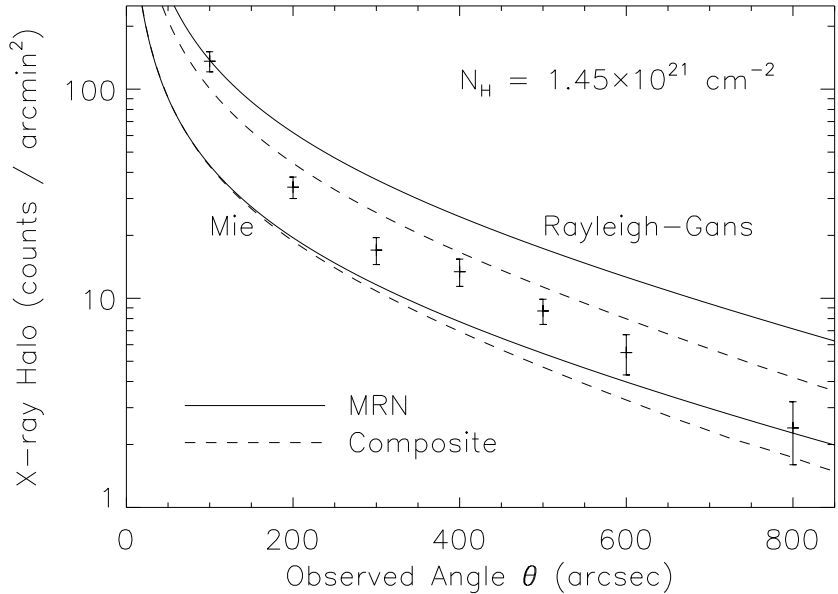}
\caption{Nova Cygni 1992 data, with MRN (bare spherical silicate and
graphite grains) and composite fluffy dust models (Mathis, 1996).  The
RG approximation overestimates the amount of scatter in both cases,
while the Mie solution underestimates the data, with similar results
for either type of dust model.  Unlike the RG approximation results,
the Mie solution does not distinguish between fluffy and dense dust.
This figure illustrates that the Nova Cygni 1992 data cannot be used
to infer the morphology of the scattering dust.\label{fig:NCwithMRN}}
\end{figure}
Our results are shown in Figure~\ref{fig:NCwithMRN}, which shows
observed halo intensity as well as the intensity calculated using the
RG approximation and the Mie solution, for both the fluffy and the
bare-core grain models.  For both types of grains, the RG
approximation overestimates the amount of scattering; in this case,
neither the MRN or the Mathis\markcite{Mathis96} (1996) composite dust
fits the data.  However, both models, when calculated using the Mie
solution, predict similar amounts of scattering that are slightly less
than the observed scattering.  While none of the four models fits the
data particularly well, we can make a number of useful observations:
\begin{enumerate}
\item The fit depends on \NH.  The measured column density was $\NH =
1.45\pm0.35\times10^{21}$\,cm$^{-2}$.  Figure~\ref{fig:NCwithMRN} uses
the central value, but using the upper $1\sigma$\ value would mean the
Mie solution models fit the data well.  To fit the RG approximation
with fluffy dust requires the lower $1\sigma$\ value, and an even
larger value for the dense dust. 
\item While there is a large difference between the fluffy and dense
dust models in the RG solution, the difference is much smaller when
using the Mie calculations.  The likely reason for this is that for a
given dust size, absorption is more important for the dense dust than
for the fluffy dust, and as a result the scattering from dense dust is
reduced relative to fluffy dust. Therefore, we conclude that the Nova
Cygni 1992 data does not distinguish between fluffy and dense dust.
\end{enumerate}

Finally, it is important to note that while the Mie solution models
marginally fit the data at larger angles, they substantially
underestimate the observed scattering at $100''$.  At small observed
angles $\theta$, dust along the LOS in the vicinity of the source
dominates the observed intensity.  Therefore, the larger halo
intensity seen at small angles may be due to an increased number of
dust grains near the source.

\begin{table}[tb]
\caption{Nova Cygni 1992 Scattering Intensity: Data (from Mathis \etal\
(1995)) and Models\label{NovaCygScat} }
\begin{tabular}{llllll} 
\hline \hline 
$\theta$ (arcsec) & Observed & MRN RG & MRN Mie & Composite RG &
Composite Mie  \\ \hline
0 & $6.62 \times 10^4$ &        &       &       &      \\
100 & $136 \pm 15$     & 137.7  & 43.2  & 100.5 & 42.9 \\
200 & $34 \pm 4$       &  62.0  & 19.5  &  44.2 & 18.7 \\ 
300 & $17 \pm 2.5$     &  36.9  & 11.6  &  25.7 & 10.9 \\
400 & $13.4 \pm 2.0$   &  24.6  &  7.73 &  16.6 & 6.94 \\
500 & $8.7 \pm 1.2$    &  17.3  &  5.45 &  11.3 & 4.68 \\
600 & $5.5 \pm 1.2$    &  12.6  &  3.98 &  7.99 & 3.26 \\
800 & $2.4 \pm 0.8$    &   7.14 &  2.27 &  4.19 & 1.73 \\ \hline
\end{tabular}
\end{table}

\section{Conclusions}

X-ray scattering is a powerful tool for measuring the composition and
size distribution of interstellar dust.  However, many approximations
and assumptions are necessary to create models of the data.  We have
assumed, for example, that dust grains are spherical, homogeneous, and
evenly distributed throughout the ISM, assumptions whose effects
should be tested.  We have explored the use of the Rayleigh-Gans
approximation for the differential scattering cross section, which
when combined with the Drude approximation to the index of refraction
provides a very convenient analytical form for discussing x-ray
scattering.  We find that the RG approximation works well for normal
interstellar dust parameters and energies at or above 2 keV, but below
this energy it overestimates the halo intensity.

We re-examined the Nova Cygni 1992 data, which has a strong x-ray
halo, and found that it does not distinguish between fluffy and dense
dust.  Mie calculations using measured optical constants produce
equally good fits for the standard MRN model, and for fluffy dust
particles.  Consequently the Nova Cygni data cannot be used as
evidence that interstellar dust particles are fluffy, though they may be.

The future of x-ray scattering is bright.  Although the ROSAT PSPC has
the highest energy-resolved spatial resolution available currently,
the AXAF CCD Imaging Spectrometer (ACIS) and High-Resolution Camera
(HRC) detectors will have arcsecond resolution with low to moderate
energy resolution.  Clearly, nearly every point source seen with AXAF
will include an x-ray halo of some kind.  The high angular resolution
will allow observers to extract halos down to $10''$\ or less, where
the scattered intensity is due almost entirely to grains very near the
source.  This will be a bonanza for dust grain theorists, as it will
provide a new handle on the size distribution of larger grains as well
as the position-dependence of grain populations.  At the same time it
will add a difficulty to the data analysis of diffuse sources, since a
calculation of the x-ray halo due to dust scattering will have to made
in order to remove it, in a process similar to the ``dereddening''
required of optical and UV measurements.

\section{Acknowledgments}

We thank Alain L\'eger for invoking our interest in this problem.
This work was performed while one of the authors (RKS) held a National
Research Council--GSFC Research Associateship.  This research was also
supported by NASA Astrophysics Theory Program grant \#344-02-05-01.

\appendix
\section{Deriving the Halo Intensity}

Consider a dust particle of size $a$\ at a point P, illuminated by an
x-ray source (S) a distance $R_1$\ away, which scatters photons (of
energy $E$) towards an observer (O) at a distance $R_2$, as is shown
in Figure~\ref{fig:scat_diag}.  Given a source luminosity of $L(E)$,
the flux at P is $L(E) \exp(-\NH(\hbox{\sc sp}) \sigma(E)) / 4\pi
R_1^2$, where $\NH(\hbox{\sc sp})$\ is the column density along the
line {\sc sp} and $\sigma(E)$\ is the average interstellar absorption
cross section at energy $E$.  Then the luminosity of scattered photons
into solid angle $d\Omega$\ is
\begin{equation}
L_{\hbox{scat}} = {{L \exp\{-\sigma(E)\NH(\hbox{\sc sp})\}}\over{4\pi R_1^2}} 
\Big({{d\sigma}\over{d\Omega}}\Big) d\Omega.
\end{equation}
Now consider the photons arriving at the observer, who measures the
flux of photons from the scattered dust particle, using a detector of
size $dA'$, and a telescope with opening angle $d\Omega'$.  Then we
can choose $d\Omega$\ so that $dA' = R_2^2 d\Omega$, and the observer
obtains a photon count rate (with the opening angle $d\Omega'$ for the
telescope) of
\begin{equation}
C = {{L \exp\{-\sigma(E) \NH(\hbox{\sc sp})\}}\over{4\pi R_1^2}}
\Big({{d\sigma}\over{d\Omega}}\Big) \exp\{-\sigma(E) \NH(\hbox{\sc po})
\} {{dA'}\over{R_2^2}} 
\end{equation}

This is the result for a single dust particle; we can now integrate it
over all dust particles in the scattering volume.  Assume that there
are $n_g(P)$\ dust grains per unit volume at position P.  We will
integrate over the line of sight using a differential volume element
with length $dR_2$ and solid angle $d\Omega'$.  The volume is $dV =
R_2^2 dR_2 d\Omega'$, and we obtain the total count rate by
integrating over $R_2$:
\begin{eqnarray}
\nonumber C_{tot} & = & \int {{L \exp\{-\sigma(E) \NH(\hbox{\sc sp})
\}}\over{4\pi R_1^2}} \Big({{d\sigma}\over{d\Omega}}\Big) n_g(P)
\exp\{-\sigma(E) \NH(\hbox{\sc po})\} {{dA'}\over{R_2^2}} R_2^2 dl 
d\Omega' \\ \nonumber
  & = & dA' d\Omega' \int {{n_g(P) L \exp(-\{\sigma(E)[\NH(\hbox{\sc
sp})+\NH(\hbox{\sc po})]\}}\over{4\pi
R_1^2}} \Big({{d\sigma}\over{d\Omega}}\Big) dR_2.
\end{eqnarray}
Thus the count rate (in photons per second) in a detector with area
$dA'$\ and opening angle $d\Omega'$ is given by $C_{tot}$.  We can
therefore write the intensity of the scattered light as
\begin{equation}
I_{sca}(E, a, \theta) = \int {{n_g(P) L \exp\{-\sigma(E)[\NH(\hbox{\sc
sp})+\NH(\hbox{\sc po})]\sigma(E)\}}
\over{4\pi R_1^2}} \Big({{d\sigma}\over{d\Omega}}\Big) dR_2. 
\end{equation}
This equation must then be integrated over a energy bandpass and a
range of dust size and position distributions.  In general, the
function $n_g$\ will be a function of dust radius $a$ as well as a
function of position.  Therefore, the total scattering for a given
energy range $\{E_1,E_2\}$\ will be
\begin{equation}
\label{Gen_Scat}
I_{sca}(\theta) = \int_{E_1}^{E_2} \int_{a_{min}}^{a_{max}}
\int_0^D {{n_g(a, P) L(E) \exp\{-\sigma(E)[\NH(\hbox{\sc
sp})+\NH(\hbox{\sc po})]\}} \over{4\pi R_1^2}} 
\Big({{d\sigma}\over{d\Omega}}\Big) dR_2\,da\,dE.
\end{equation}

This equation is completely general; however, we can simplify it
substantially if the observed angle $\theta$\ is very small, limiting
the observations to very small angles.  In this case, the path
traveled by a scattered photon will be very close to that traveled by
an unscattered photon and the total hydrogen column density for the
source, $\NH(\hbox{\sc so})$, will be nearly the same as that of the
scattered photons, $\NH(\hbox{\sc sp}) + \NH(\hbox{\sc po})$.  As a
result, we have the following approximation:
\begin{equation}
{{L(E) \exp\{-\sigma(E)[\NH(\hbox{\sc sp})+\NH(\hbox{\sc
po})]\} }\over{4\pi R_1^2}} \approx {{F_X(E)
\cos(\phi-\theta)}\over{(1-z)^2}} 
\end{equation}
where $F_X(E)$\ is the observed flux from the source.  We can then
rewrite equation (\ref{Gen_Scat}) in terms of the normalized projected
position coordinate $z$\ and the scattering angles $\theta$\ and
$\phi$:
\begin{equation}
\label{First_Approx}
I_{sca}(E_{b}, \theta) = \int_{E_b} \int_{a_{min}}^{a_{max}}
\int_0^1 {n_g(a, z,\theta) {F_X(E)
\cos(\phi-\theta)} \over{(1-z)^2}}
\Big({{d\sigma}\over{d\Omega}}\Big) {{D}\over{\cos\theta}}\,dz\,da\,dE
\end{equation}

We can reduce this further by assuming that the dust size and position
distribution is separable, and is independent of the hydrogen density
$n_H$, so that
\begin{equation}
n_g(a, z, \theta) da = n(a) f(z) n_H(z) da.
\end{equation}
We have used the assumption of similar paths for the scattered and
direct photons here as well.  Using these, and assuming $\cos\theta
\approx 1$, we can rewrite (\ref{First_Approx}), to get
\begin{equation}
I_{sca}(\theta) = \NH \int_{E_1}^{E_2} F_X(E)
\int_{a_{min}}^{a_{max}} n(a) \int_0^1 {f(z) { \cos(\phi-\theta)}
\over{(1-z)^2}} \Big({{d\sigma(E, a, \phi)}
\over{d\Omega}}\Big)\,dz\,da\,dE  
\end{equation}
where $\NH \equiv \int n_H(z) D dz$\ is the hydrogen column density to
the source.  The final assumption necessary to derive equation
(\ref{scat_eq}) in the text is that $\cos(\phi-\theta)
\approx 1$.  We have already limited the observed angle $\theta$\
to small angles.  Although close to the x-ray source $\phi$\ can be
large even for very small values of $\theta$, $(d\sigma/d\Omega)$\ is
strongly suppressed at scattering angles above $1^\circ$, so that in
practice, all the contribution to the scattered light comes from
scattering angles below $1^\circ$.  Therefore the actual scattering
angle $\phi$\ will be small, so $\cos(\phi-\theta) \approx 1$.  Note
also that $\phi > \theta$\ for all geometries.  For a nonuniform dust
distribution, in which a significant amount of dust is near the source
the contribution from larger scattering angles may be important, and
an exact expression for $\cos(\phi-\theta)$\ will then be required.

\end{document}